\documentclass[aps,pra,twocolumn,10pt,showpacs]{revtex4}

\long\def\/*#1*/{}
\usepackage{amsfonts}
\pdfoutput=1
\usepackage{graphicx}
\usepackage{amsmath}
\usepackage{amsthm}
\usepackage[colorlinks=true, citecolor=blue, urlcolor=blue ]{hyperref}
\usepackage{graphicx}

\begin{document}

\title{Study of Quantum Walk over a Square Lattice }

\author{Arkaprabha Ghosal}
\email{a.ghosal1993@gmail.com }
\affiliation{Department of Physics and Center for Astroparticle Physics and Space Science, Bose Institute, Bidhan Nagar
Kolkata - 700091, India.}

\/*
\author{Somshubhro Bandyopadhyay}
\email{som.s.bandyopadhyay@gmail.com }
\affiliation{Department of Physics and Center for Astroparticle Physics and Space Science, Bose Institute, Bidhan Nagar
Kolkata - 700091, India.}
*/
\author{Prasenjit Deb}
\email{devprasen@gmail.com}
\affiliation{Department of Physics and Center for Astroparticle Physics and Space Science, Bose Institute, Bidhan Nagar
Kolkata - 700091, India.}

\begin{abstract}
Quantum random walk finds application in efficient quantum algorithms as well as in quantum network theory. Here we study the mixing time of a discrete quantum walk over a square lattice in presence of percolation and decoherence. We consider bit-flip and phase damping noise, and evaluate the instantaneous mixing time for both the cases. Using numerical analysis we show that in the case of phase damping noise probability distribution of walker's position is sufficiently close to the uniform distribution after infinite time. However, during the action of bit-flip noise, even after infinite time the total variational distance between the two probability distributions is large enough.
        
\end{abstract}

\maketitle


\section{Introduction}
Quantum mechanics has puzzled the scientists since its inception due to the counter-intuitive characteristics.
Recent experimental and theoretical developments in quantum information and computation have revealed a lot of important outcomes which highlight the advantages of using quantum mechanical systems and resources in performing various tasks. Different fascinating effects have been uncovered which are strikingly different from their classical counterparts, both from the physical point of view as well as from a computer science and communication theory perspective. Quantum teleportation\cite{PhysRevLett.70.1895}, dense coding\cite{PhysRevLett.69.2881}, state merging\cite{nature03909}, quantum cryptography\cite{PhysRevLett.67.661,PhysRevLett.68.557}, quantum computation\cite{Nielsen:2011:QCQ:1972505} etc are possible \emph{iff} quantum mechanical resources are available. In the classical world neither any such phenomena will take place nor any such task can be accomplished. So, in order to exploit the quantum effects modern technologies are being developed which in turn will enable us to make use of quantum resources on a large scale.

\paragraph*{}
In classical computer science random walks play a vital role\cite{Motwani:1995:RA:211390}. So it is plausible to think that the quantum counterpart will be equally important in the study of quantum computation, and infact, recent research confirms that the notion of \emph{quantum random walk} has emerged as an important element in the development of quantum algorithms\cite{PhysRevA.67.052307,Childs:2003:EAS:780542.780552}. A lot of algorithms have been developed using quantum walks, which display speed-ups compared to their classical equivalents. One such algorithm is the algorithm for element distinctness\cite{Ambainis:2007:QWA:1328722.1328730} where quantum walks appear dramatically to make it the most efficient.

\paragraph*{}
There are two natural variants of basic model of quantum random walk. One is the \emph{continuous} model and another one is the \emph{discrete} model. The continuous model was introduced by Childs \emph{et. al}\cite{Childs2002} and the discrete model by Aharonov \emph{et. al}\cite{Aharonov:2001:QWG:380752.380758}. Though the discrete and continous time quantum walks have different origin, they can be precisely related to each other. The dynamics of quantum random walks have been analyzed for example, on the line\cite{ROMANELLI2004395}, circle\cite{PhysRevA.73.012313}, hyperlattice\cite{PhysRevA.74.020102} and hypercube\cite{Moore} and the theoretical properties of quantum walks on general graphs have been outlined in Ref.\cite{doi:10.1142/S0219749906002195} Some important properties of quantum random walk on directed graphs have been provided in Ref.\cite{Montanaro:2007:QWD:2011706.2011711}.

\paragraph*{}
Ideally, quantum systems are taken to be isolated from the environment and the evolution of such systems are described by \emph{unitary} evolution. But, in the real scenario the quantum systems are susceptible to imperfections and interactions with their environment, due to which \emph{decoherence} takes place. Sufficient amount of decoherence can remove any potential benefits from quantum dynamics. Therefore, for practical implementation of quantum protocols in communication or computation it is necessary to study the robustness of such walks in presence of decoherence.

\paragraph*{}
Decoherence is unarguably a nemesis of quantum information processing. However, in the context of quantum walks it can infact be useful\cite{PhysRevA.67.042315}. For example, decoherence can be used to force a quantum walk to move towards a steady state distribution, which has been nicely illustrated for hypercube\cite{Drezgich:2009:CCM:2011804.2011812}, line\cite{PhysRevA.67.042315} and N-cycles\cite{1367-2630-9-4-087}.
Elegant analysis has been done by Richter\cite{1367-2630-9-3-072,PhysRevA.76.042306} showing the quantum speedups of classical mixing processes under a restricted decoherence model. By examining the discrete-time quantum walk on hypercube Marquezino \emph{et. al}\cite{PhysRevA.77.042312} derived the limiting time-averaged distribution in the coherent case. It has been also shown that the walk approaches the uniform distribution in the decoherent case and there is an optimal decoherence rate which provides the fastest convergence\cite{PhysRevA.77.042312}. Though the above mentioned literatures show the effect of decoherence on quantum walks, more study is undoubtedly necessary.

\paragraph*{}
In this article we aim to investigate whether decoherence always forces the quantum walk to move towards uniform distribution or not by analysing the mixing time. For our purpose we consider discrete time quantum walk on a square lattice and assume that the graph structure changes randomly, i.e., the walk is taking place on a dynamical percolation graph. The walker is a qubit (two-state quantum system) and the decohering noises are bit-flip and phase damping noise. We find that in presence of decoherence quantum walk does not always reach the uniform distribution. To be more specific, in case of phase damping noise the probability distribution of walker's position is sufficiently close to the uniform distribution after infinite time, whereas, the walk can not reach the uniform distribution under the action of bit-flip noise, even after infinite time. The article is arranged as follows. In Sec [II] we provide an overview of quantum random walk on dynamical percolation graphs. The main results are  presented in Sec[III]. Conclusion about our work is provided in Sec[IV].


\section{Discrete Quantum Random walk on dynamical percolation graphs}
A discrete-time quantum walk (DTQW) is modeled as a particle consisting of a two-level coin (a qubit) existing in the Hilbert space $\mathcal{H}_c$ and a position degree of freedom existing in the Hilbert space $\mathcal{H}_p$. The Hilbert space of the quantum walk is therefore the tensor product of $\mathcal{H}_c$ and $\mathcal{H}_p$, i.e., $\mathcal{H}= \mathcal{H}_c\otimes \mathcal{H}_p$. The coin space $\mathcal{H}_c$ is spanned by the basis states of the coin and the position space $\mathcal{H}_p$ is spanned by the orthonormal basis vectors corresponding to the vertices of the graph on which the quantum walk occurs. The unitary operator that defines the time evolution of DTQW and acts on the Hilbert space $\mathcal{H}$ is given by
\begin{equation}
U=S\cdot (C\otimes I)
\end{equation}
where $S$, $C$, $I$ denote shift operator, coin operator and identity operator respectively. DTQW has been rigorously studied for different graphs assuming that the underlying graph does not change with time. However, the underlying graph structure can be changed and it can be done by keeping an edge with some probability, say, $\lambda$ or discarding it with probability $1-\lambda$. In graph theory this type of  structural change is termed as \emph{percolation}. If the graph changes randomly then the percolation is termed as \emph{dynamical percolation}. The unitary evolution of the quantum walk is disturbed by dynamical percolation, leading to a special type of open system causing decoherence, which can be viewed also as a source of noise. 
\paragraph*{}
From the definition of percolation it is clear that different graph structures mean different edge configurations and the time evolution of quantum walk on each configuration is described by a unitary operator. The unitary operator that describes the time evolution of a quantum walk on a percolated graph for edge configuration $\mathcal{K}$ is defined as
\begin{equation}\label{unitary}
U_{\mathcal{K}}=S_{\mathcal{K}}\cdot (C\otimes I)
\end{equation}
where $S_{\mathcal{K}}$ is the shift operator for the configuration $\mathcal{K}$. The dynamic percolation is introduced in this process (quantum walk) by choosing different edge configurations $\mathcal{K}$ at each step, and thus random application of different unitary step operators. The state of the walker is best described by density matrix. Now if we denote the probability of choosing a configuration $\mathcal{K}$ and corresponding unitary step $U_{\mathcal{K}}$ by $\pi_{\mathcal{K}}$ then the time evolution of the walker during the percolated quantum walk is given by
\begin{equation}
\Phi(\rho)\equiv \sum_{\mathcal{K}=1}^{R}\pi_{\mathcal{K}}U_{\mathcal{K}}\rho U_{\mathcal{K}}^{\dagger}
\end{equation}
where $\Phi(\cdot)$ is a \emph{completely positive trace preserving} (CPTP) map and $R$ is the total number of  edge configurations. The shift operator $S_{\mathcal{K}}$ for configuration $\mathcal{K}$ depends on the graph structure. However, the coin operator $C$ is independent of the graph structure, and it can be expressed as a function of two angles $\alpha$ and $\beta$. The coin operator looks like
\begin{equation}
C(\alpha,\beta) =  \begin{pmatrix} \mbox{i}e^{-\mbox{i}\alpha}sin\beta& cos\beta \\cos\beta&\mbox{i}e^{\mbox{i}\alpha} sin\beta\end{pmatrix}
\end{equation}
 where `i' has its usual meaning. Inserting $\alpha=\frac{\pi}{2}$ and $\beta=\frac{\pi}{4}$ we get the Hadamard operator.
 \paragraph*{}
 Along with percolation if the walker (qubit) faces Markovian noise then the time evolution of the qubit can be given by
 \begin{equation}
 \Phi(\rho)=\sum_{\mathcal{K}}~\pi_{\mathcal{K}}\{(1-p)U_{\mathcal{K}}\rho U_{\mathcal{K}}^{\dagger} +p\sum_m~U_{\mathcal{K}}M_m\rho M_m^{\dagger}U_{\mathcal{K}}^{\dagger}\}
 \end{equation}
 where $M_m$ are the \emph{Kraus operators} satisfying the condition $\sum_m~M_m^{\dagger}M_m=1$, and $m$ is the \emph{Kraus rank}. Here $p$ is the probability with which noise acts on the qubit. Clearly, the above equation describes discrete-time quantum walk on a dynamical percolation graph in presence of external noise.
 
\section{Quantum Walk over a square lattice}
 We consider a discrete-time quantum walk over a square lattice and assume that dynamic percolation is occuring during the walk. Moreover, Markovian noise is acting on the walker, which is a qubit here. We analyse the evolution of the walker's state and study the mixing time by constructing zone basis.
\paragraph*{}
The zone is a boundary starting from the nearest neighour of any choice
of vertex, increasing in size as we move far from our chosen vertex.
The number of vertex or nodes which fall on a zone boundary depends
on the coordination number $C$. In case of square lattice one have
$C=4$, for triangular lattice $C=6$ and so on. In case of square
lattice, the largest boundary will contain $\frac{N}{2}$ number of nodes, where $N$= total number of vertices in the lattice.
Thus, for a square lattice number of nodes on a zone boundary increases as a multiple
of 4. To construct zone basis we start from the origin and define the basis vectors as
\begin{eqnarray}
\lvert Z_{0}\rangle &=&\lvert0,0\rangle \nonumber\\
\lvert Z_{1}\rangle&=&\lvert0,1\rangle +\lvert1,0\rangle 
+\lvert0,-1\rangle +\lvert-1,0\rangle \nonumber\\
\lvert Z_{2}\rangle &=&\lvert2,0\rangle +\lvert1,1\rangle +\lvert0,2\rangle +\lvert-1,1\rangle\nonumber\\ &+&\lvert-2,0\rangle +\lvert-1,-1\rangle +\lvert0,-2\rangle +\lvert1,-1\rangle 
\end{eqnarray}
and so on. The zone basis $\left|Z_{M}\right\rangle $ corresponding to the boundary of the $M^{th}$ zone is a linear combination of $4M$ no. of vertex states. Having defined the zone basis we now find out the shift operators which control the random walk. As we have considered square lattice, two shift operators are needed to describe the walk.
For $\mathcal{K}^{th}$ realization of graph structure these two operators are given by\cite{1751-8121-46-10-105306}
\begin{eqnarray}
S_{\mathcal{K}}^{x}&=&\sum_{\{x,y\}}\sum_{c=0}^{1}(\sum_{\{(X,y),(x,y)\}\in\kappa}\lvert c\rangle\langle c\lvert\otimes\lvert X,y\rangle \langle x,y\lvert \nonumber\\
&+&\sum_{\{(X,y),(x,y)\}\notin\kappa}(\sigma_{x}\otimes I)\lvert c\rangle \langle c\lvert \otimes\lvert x,y\rangle \langle x,y\lvert)\nonumber\\
S_{\mathcal{K}}^{y}&=&\sum_{\{x,y\}}\sum_{c=0}^{1}(\sum_{\{(x,Y),(x,y)\}\in\kappa}\lvert c\rangle \langle c\lvert\otimes\lvert x,Y\rangle \langle x,y\lvert \nonumber\\
&+&\sum_{\{(x,Y),(x,y)\}\notin\kappa}(\sigma_{x}\otimes I)\lvert c\rangle\langle c\lvert \otimes\lvert x,y\rangle \langle x,y\lvert)
\end{eqnarray}
where $X=x\oplus c$ and $Y=y\oplus c$.
The unitary operator defined in Eq.(\ref{unitary}) can, therefore, be written as
$U_{\kappa}^{(x,y)}=\{S_{\kappa}^{y}.(C_{(\alpha,\beta)}\otimes I)\}\{S_{\kappa}^{x}.(C_{(\alpha,\beta)}\otimes I)\}$. The shift operators determine the evolution of the walker's position with time.
However, in case of an empty graph the walker has no option to shift anywhere, only the spin state will change. Therefore, for an empty graph the resulting effect on the walker can be expressed as 
$\xi_{\kappa=0}(\rho)=(\sigma_{x}C_{(\alpha,\beta)}\otimes I)^{2}\rho\{(\sigma_{x}C_{(\alpha,\beta)}\otimes I)^{2}\}^{\dagger}$
 where $\mathcal{K}=0$ denotes empty graph realization. Thus by choosing an empty graph
one can easily discard vertex system $\mathcal{H^{N}}$. If Markoian
noise simultaneously affects walker then operation on $\mathcal{H}^{2}$
will be \cite{kendon_2007,Novotny}
\begin{eqnarray}
\xi_{0}(\rho^{c})&=&(1-p)(\sigma_{x}C_{(\alpha,\beta)})^{2}\rho^{c}\{(\sigma_{x}C_{(\alpha,\beta)})^{2}\}^{\dagger}\nonumber\\
&+&p\sum_{l}(\sigma_{x}C_{(\alpha,\beta)})^{2}K_{l}\rho^{c}K_{l}^{\dagger}\{(\sigma_{x}C_{(\alpha,\beta)})^{2}\}^{\dagger}
\end{eqnarray}
where, $\rho^{c}\in\mathcal{L}(\mathcal{H}^{2})$, $\xi_{0}:\mathcal{L}(\mathcal{H}^{2})\rightarrow\mathcal{L}(\mathcal{H}^{2})$
is the mapping and $K_l$ are the Krauss operators satisfying the condition $\sum_l~K_l^{\dagger}K_l=1$.
The superscript $c$ stands for coin state. Since empty graph operation acts only on the spin state, we can use it
to construct the walker's state after arbitrary iteration. For that purpose, we first need to find out the position probabilities of the walker in different zones. Starting from the origin the walker will traverse the entire lattice through succesive iterations, and hence we can define the position
probabilities of the walker in different zones as 
\begin{eqnarray}
P_{0}&=\mbox{Tr}[(I\otimes\left|Z_{0}\right\rangle \left\langle Z_{0}\right|)\rho]\nonumber\\
P_{1}&=\frac{1}{4}\mbox{Tr}[(I\otimes\left|Z_{1}\right\rangle \left\langle Z_{1}\right|)\rho]\nonumber\\
P_{2}&=\frac{1}{8}\mbox{Tr}[(I\otimes\left|Z_{2}\right\rangle \left\langle Z_{2}\right|)\rho]\nonumber\\
&\vdots \nonumber\\
P_{M}&=\frac{1}{4M}\mbox{Tr}[(I\otimes\left|Z_{M}\right\rangle \left\langle Z_{M}\right|)\rho]
\end{eqnarray}


Using the above defined probability distribution we can finally write the state of the walker when it reached $M^{th}$ zone as
\begin{eqnarray}
\rho&=&P_{0}\xi_{0}^{n_{e}}(\rho_{0}^{c})\otimes\left|Z_{0}\right\rangle \left\langle Z_{0}\right| \nonumber\\
&+&\sum_{m=1}^{M}P_m\rho^c_m\otimes\lvert Z_m\rangle\langle Z_m\lvert
\end{eqnarray}
where, $\rho_1^c=\lambda\xi_{0}^{n_{e}}(\rho_{0}^{c})+(1-\lambda)\xi_{0}^{n_{e}-3}(\rho_{0}^{c})$,
$\rho_2^c=\lambda\xi_{0}^{n_{e}-3}(\rho_{0}^{c})+(1-\lambda)\xi_{0}^{n_{e}-6}$, $\rho_M^c= \xi_{0}^{n_{e}-3m}(\rho_{0}^{c})$ and
$P_{0}>P_{1}>P_{2}>.........>P_M$
holds until the walker reaches the uniform distribution. From the above equation it is clear that the 
state of the walker depends explicitly on the weight of edges
$\lambda\in[0,1]$ but not on any specific realization occuring randomly.
The operation $\xi_{0}$ acts only on the coin space (qubit space) and changes the
direction of the walk. The total number of effective iterations which a qubit has undergone while being in a particular zone is given by $n_e$. In our case $n_{e}=3m+n$, where $m$ and $n$ denotes zone and iteration number respectively, $0\leq m\leq M=\frac{\sqrt{N}}{2}$ and $0\leq n\leq2$ , $N$ is the total number of vertices in the lattice. 
\/*
Now, suppose
we want to bound the walk inside $(M\times M)$ sublattice. Then we
can write our iteration number for $m<M$ where $0\leq n\leq2$ as
\[
n_{e}=3m+n
\]
and for $m=M$ where $3M\leq n\leq\infty$as
\[
n_{e}=n
\]
Thus, at every $3^{rd}$ step of iteration, a convex set is formed
at each vertex of the increasing zone boundary. The convex set here
implies that if one or more edges fall between any two consecutive
zone, there spin state will correlate in the same direction. Otherwise,
the direction will correlate with its next larger zone. 
*/
\paragraph*{}
Now, depending on the different values of the percolation parameter $\lambda$ the graph structure changes in several ways. For $\lambda=\frac{1}{2}$, edges will fall or break randomly, and therefore, the walk will be fully
unbiased if one chooses Hadamard coin. When $\lambda\rightarrow0$,
the spreading of walk will be much much slower. For $\lambda=1$, there
will be perfect anti-correlation between spin states situated at two
antipodal points of any particular zone. When $\lambda<1$ this perfect
anti-correlation is lost due to percolation. 
\/*
Therefore, the walker's state
$\rho(n,m,\lambda)$ will change as 
\begin{eqnarray}
\rho(0,0,\lambda)&=&P_{0}(0,0)\xi_{0}^{0}(\rho_{0}^{c})\otimes\left|Z_{0}\right\rangle \left\langle Z_{0}\right|\nonumber\\
&=&\rho_{0}^{c}\otimes\left|Z_{0}\right\rangle \left\langle Z_{0}\right|\nonumber\\
\rho(1,0,\lambda)&=&\xi_{0}^{1}(\rho_{0}^{c})\otimes(P_{0}(1,0)\left|Z_{0}\right\rangle \left\langle Z_{0}\right|+P_{1}(1,0)\left|Z_{1}\right\rangle \left\langle Z_{1}\right|)\nonumber\\
\rho(2,0,\lambda)&=&\xi_{0}^{2}(\rho_{0}^{c})\otimes(P_{0}(2,0)\left|Z_{0}\right\rangle \left\langle Z_{0}\right|+P_{1}(2,0)\left|Z_{1}\right\rangle \left\langle Z_{1}\right|)\nonumber\\
\rho(0,1,\lambda)&=&P_{0}(0,1)\xi_{0}^{3}(\rho_{0}^{c})\otimes\left|Z_{0}\right\rangle \left\langle Z_{0}\right|+P_{1}(0,1)\{\lambda\xi_{0}^{3}(\rho_{0}^{c})\nonumber\\
&+&(1-\lambda)\xi^{0}(\rho_{0}^{c})\}\otimes\left|Z_{1}\right\rangle \left\langle Z_{1}\right| \nonumber\\
&+& P_{2}(0,1)\left|Z_{2}\right\rangle \left\langle Z_{2}\right|\otimes\xi_{0}^{0}(\rho_{0}^{c})
\end{eqnarray}
and so on.
*/
For our purpose we choose the limit for the percolation parameter as $0<\lambda\leq\frac{1}{2}$.
The reason behind this choice is that in this limit the walker can not find any large connected
path between two widely separated zones.
Therefore, upto two steps of iteration the walk will be bounded
between two consecutive zones. 

\paragraph*{}
In the next we will analyse the mixing time of the quantum walk 
considering two types of noises: continuous phase damping and bit-flip.

\subsection{Percolation with continuous phase damping noise}
Let us consider that phase damping noise is acting continuously on the walker (qubit) and the noise parameter $p=\Gamma_{dep}dt$, where $\Gamma_{dep}$ denotes dephasing rate. For continuous dephasing $(1-p_{dep})^{n}=e^{-\Gamma_{dep}t}$. Now, if we choose the initial state of the walker as 
\begin{equation}
\rho_{0}^{c}=\{a\left|\psi_{\theta,\phi}\right\rangle \left\langle \psi_{\theta,\phi}\right|+(1-a)\frac{I_{2}}{2}\}\otimes\left|0,0\right\rangle \left\langle 0,0\right|
\end{equation}
then the empty graph operation looks like
\begin{eqnarray}
\xi_{0}(\rho_{0}^{c})&=&\frac{1}{2}\begin{pmatrix} 1-a\mbox{cos}\theta & a\mbox{sin}\theta e^{-i\phi}(1-p_{dep})^n\\ a\mbox{sin}\theta e^{i\phi}(1-p_{dep})^n & 1+a\mbox{cos}\theta
\end{pmatrix}\nonumber\\
&=&\frac{1}{2}\begin{pmatrix}1-a\mbox{cos}\theta & a\mbox{sin}\theta e^{-i\phi}e^{-\Gamma_{dep}t}\\
a\mbox{sin}\theta e^{i\phi}e^{-\Gamma_{dep}t} & 1+a\mbox{cos}\theta \end{pmatrix}\nonumber
\end{eqnarray}
Therefore, the state of the walker will be
\begin{eqnarray}
\rho_{dep}(t)&=&P_{0}\xi_{0}^{n_{e}}(\rho_{0}^{c})\otimes\left|Z_{0}\right\rangle \left\langle Z_{0}\right|\nonumber\\
&+&P_{1}\rho_1^c\otimes\left|Z_{1}\right\rangle \left\langle Z_{1}\right|\nonumber\\
&+&P_{2}\rho_2^c\otimes\left|Z_{2}\right\rangle \left\langle Z_{2}\right|\nonumber\\
&+&..........\nonumber\\
&+&P_{M}\rho_n^c\otimes\left|Z_{M}\right\rangle \left\langle Z_{M}\right|\nonumber\\
\end{eqnarray}
where, $\rho_1^c=\lambda\xi_{0}^{n_{e}}(\rho_{0}^{c})+(1-\lambda)\xi_{0}^{n_{e}-3}(\rho_{0}^{c})$,
$\rho_2^c=\lambda\xi_{0}^{n_{e}-3}(\rho_{0}^{c})+(1-\lambda)\xi_{0}^{n_{e}-6}$, $\rho_M^c= \xi_{0}^{n_{e}-3m}(\rho_{0}^{c})$. The condition $\mbox{Tr}[\rho_{dep}]=1$ implies that the walk is bounded within $M^{th}$ zone. 
\paragraph*{}
For an $N$ vertex lattice, the trace norm between $\rho_{dep}(t)$ and the final asymptotic state $\rho_{dep}(\infty)$\cite{PhysRevLett.108.230505} is given by
\begin{equation}
D[\rho_{dep}(t),\rho_{dep}(\infty)]=\frac{1}{2}||\rho_{dep}(t)-\rho_{dep}(\infty)||_{1}
\end{equation}
If we define, $\rho^c(t)=\mbox{Tr}_{p}[\rho(t)]$ then it can be shown that  
\begin{equation}
D[\rho_{dep}(t),\rho_{dep}(\infty)]\geq D[\rho_{dep}^c(t),\rho_{dep}^c(\infty)]
\end{equation}
Thus for an $N$ vertex square lattice, trace norm can be calculated as
\begin{eqnarray}
2D[\rho_{dep}^c (t),\rho_{dep}^c(\infty)&\leq&||P_{0}\xi_{0}^{n_{e}}(\rho_{0}^{c})-\frac{I_{2}}{2N}||_{1}\nonumber\\
&+&4||P_{1}\xi_{0}^{n_{e}-3}(\rho_{0}^{c})-\frac{I_{2}}{2N}||_{1}\nonumber\\
&+&8||P_{2}\xi_{0}^{n_{e}-6}(\rho_{0}^{c})-\frac{I_{2}}{2N}||_{1}\nonumber\\
&+&........\nonumber\\
&+&4M||P_{M}\xi_{0}^{n_{e}-3M}(\rho_{0}^{c})-\frac{I_{2}}{2N}||_{1}\nonumber\\
\end{eqnarray}
Due to percolation and the dephasing noise the coin state will become maximally mixed for $n_{e}=(3m+n)\rightarrow\infty$. Now, position probability distributions of the walker at each zone will be nearly equal at mixing time $(t\simeq t_{mix}<\infty)$. To study the asymptotic behaviour we choose $N\simeq\mathcal{O}(10^{4})$ or higher. For $t\simeq t_{mix}$ one can say 
\begin{eqnarray}
|P_{0}(t)-P_{u}|&\simeq&|P_{1}(t)-P_{u}|\nonumber\\
&\simeq&|P_{2}(t)-P_{u}|\nonumber\\
&\simeq&............\nonumber\\
&\simeq&|P_{M}(t)-P_{u}|\simeq\delta_{dep}\nonumber
\end{eqnarray} 
where, $P_{u}=\frac{1}{\left(\frac{N}{2}\right)}$ is the uniform distribution of the postion of the walker inside $(M\times M)$ sublattice. Choosing larger and larger lattice will also increase the size of this sublattice and thus we can cover a larger set. Therefore, for $t\simeq t_{mix}$ we can write
\begin{equation}
P_{0}(t)\simeq P_{1}(t)\simeq P_{2}(t)
\simeq...\simeq P_{M}(t)\simeq(\delta_{dep}+\frac{2}{N})
\end{equation}
Substituting the values of the probabilities from Eq.() in Eq.() one gets for $t=t_{mix}$,
\begin{eqnarray}
2D[\rho_{dep}^c(t),\rho_{dep}^c(\infty)]\leq||(\delta_{dep}+\frac{2}{N})\xi_{0}^{n_{e}}(\rho_{0}^{c})-\frac{I}{2N}||_{1}\nonumber\\
+4||(\delta_{dep}+\frac{2}{N})\xi_{0}^{n_{e}-3}(\rho_{0}^{c})-\frac{I}{2N}||_{1}+...\nonumber\\
+4M||(\delta_{dep}+\frac{2}{N})\xi_{0}^{n_{e}-3M}(\rho_{0}^{c})-\frac{I}{2N}||_{1}\nonumber\\
\end{eqnarray}
By putting $\lambda=\frac{1}{2}$ in the above equation we get 
\begin{equation}
2D[\rho_{dep}(t),\rho_{dep}(\infty)]\leq \delta_{dep}+4MA+4\sum_{k=1}^{M-1}kB
\end{equation}
where, $A=[\delta_{dep}^{2}+a^{2}(\delta_{dep}+\frac{2}{N})^{2}e^{-2\Gamma_{dep}t}]^{\frac{1}{2}}$,
$B=[\delta_{dep}^{2}+(\delta_{dep}+\frac{2}{N})^{2}a^{2}sin^{2}\theta cos^{2}\phi e^{-2\Gamma_{dep}t}]^{\frac{1}{2}}$. Now, by taking the approximation
$
a^{2}(1+\frac{2}{N\delta_{dep}})^{2}e^{-2\Gamma_{dep}t}\ll1
$
 and choosing $N\gg1$ we get from straight forward calculation
\begin{equation}
D[\rho_{dep}^c(t),\rho_{dep}^c(\infty)]\leq \frac{1}{4}N\delta_{dep}+a^{2}MX+a^{2}MY
\end{equation}
\\
where $X=\delta_{dep}(1+\frac{2}{N\delta_{dep}})^{2}e^{-2\Gamma_{dep}t}$ and $Y=\delta_{dep}(1+\frac{2}{N\delta_{dep}})^{2}sin^{2}\theta cos^{2}\phi\{e^{-2(t_{mix}+3)\Gamma_{dep}}\frac{(1-e^{-3M\Gamma_{dep}})}{(1-e^{-6\Gamma_{dep}})}\ $. Now, the coin state that we have considered has Bloch sphere representation. Therefore, taking average over all possible coin states we have the average trace distance as
\begin{eqnarray}
\overline{D}[\rho_{dep}^c(t),\rho_{dep}^c(\infty)]&=&\frac{1}{4\pi}\iint_{(\theta,\phi)}D(\theta,\phi)sin\theta d\theta d\phi\nonumber\\
&=&\frac{1}{4}N\delta_{dep}+a^{2}MX \nonumber\\
&&\times\left\{ 1+\frac{f(M,\Gamma_{dep})}{3}\right\} 
\end{eqnarray}
\\
where, $f(M,\Gamma_{dep})=e^{-6\Gamma_{dep}}\frac{(1-e^{-3M\Gamma_{dep}})}{(1-e^{-6\Gamma_{dep}})}$ is a function of the size of lattice and dephasing rate.  Hence, we have evaluated the trace norm nearly arround the mixing time. Now, there comes a question. Using the noise rate can we estimate trace norm for all $t<t_{mix}$, such that the trace norm can be written as\\ 
\begin{eqnarray}
\overline{D}[\rho_{dep}^c(t),\rho_{dep}^c(\infty)] &=&\left\langle D(t=\infty)\right\rangle\nonumber\\
&&+\left\langle D[\rho(t=0),\rho(N,t=\infty)]\right\rangle e^{-2\Gamma_{dep}t}\nonumber\\
&&-\left\langle D(\infty)\right\rangle e^{-2\Gamma_{dep}t}
\end{eqnarray}
We choose $\delta_{dep}=\frac{1}{2N\sqrt{N}}$ as the closeness with the uniform distribution $P_{u}$ inside $(M\times M)$ sublattice. Choosing $\delta_{dep}\simeq\mathcal{O}(N^{-\frac{3}{2}})$ makes the quantity $M\delta_{dep}\left(1+\frac{2}{N\delta_{dep}}\right)^{2}\simeq\mathcal{O}(1)$ and then we can write 
\[
a^{2}\left\{ 1+\frac{f(M,\Gamma_{dep})}{3}\right\} =\frac{1}{2}\sqrt{(1-\frac{1}{N})^{2}+a^{2}}
\]
and hence 
\begin{equation}
f(M,\Gamma_{dep})=3\left\{ \frac{\sqrt{(1-\frac{1}{N})^{2}+a^{2}}}{2a^{2}}-1\right\} 
\end{equation}
Therefore, for a fixed coin parameter $a$ and fixed vertex number $N$, we get a specific value of $f(M,\Gamma_{dep})$. Plotting $f(M,\Gamma_{dep}) ~\mbox{vs}.~ \Gamma_{dep}$ we can estimate those tunable dephasing rates for which the following equation holds in the range $0\leq t\leq\infty$:
\begin{eqnarray}
\overline{D}[\rho_{dep}^c(t)),\rho_{dep}^c(\infty)] &=& \frac{1}{4\sqrt{N}}\{1-\mbox{e}^{-2t\Gamma_{dep}}\}\nonumber\\
&&+\frac{\sqrt{\left(1-\frac{1}{N}\right)^{2}+a^{2}}}{2}\times \mbox{e}^{-2t\Gamma_{dep}t}\nonumber\\
\end{eqnarray}

\begin{figure}[t!]
\centering
\includegraphics[height=7cm,width=7cm]{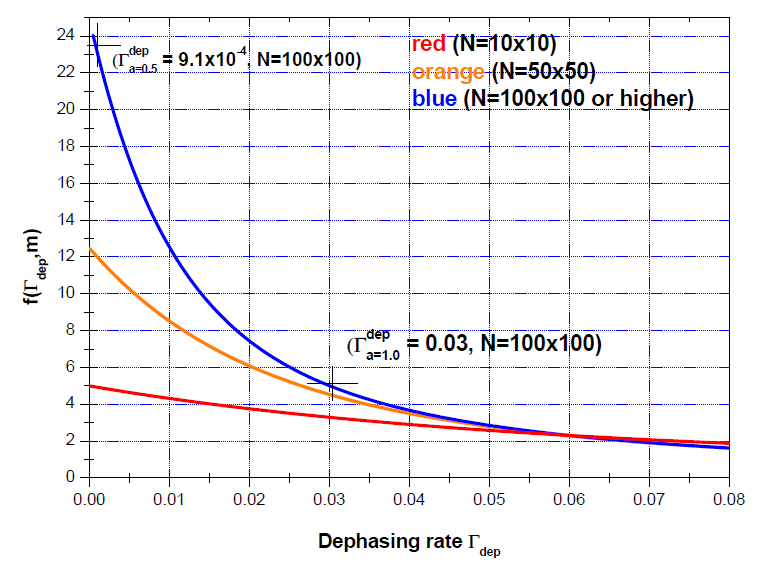}
\caption{(Color on-line) Tunable dephasing rate for $N\sim10^4$ or higher order lattice size. }\label{fig2}
\end{figure}

\begin{figure}[t!]
\centering
\includegraphics[height=7cm,width=7cm]{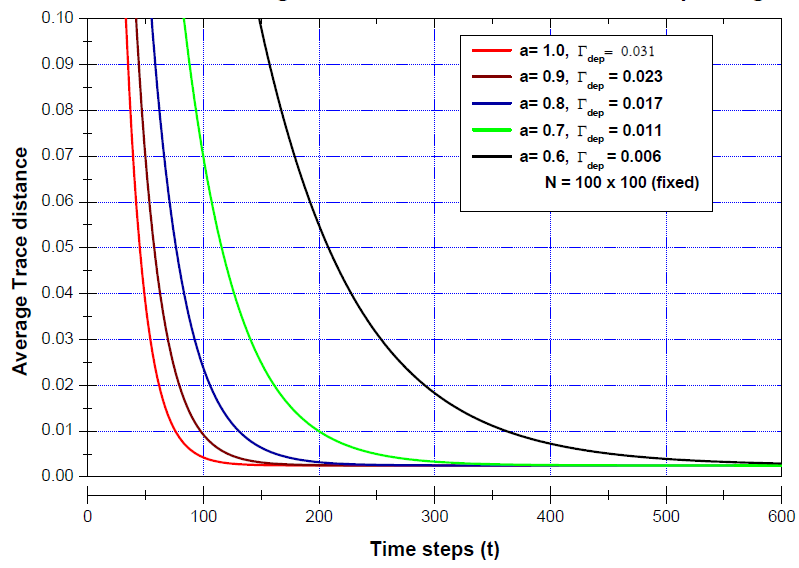}
\caption{(Color on-line) Time evolution of average trace distance ($\overline{D}(t,\delta)$) under continuous dephasing. }\label{fig3}
\end{figure}
From the above equation one can imply an interesting phenomenon. Suppose the walker has passed several zones without knowing how many different or similar realizations has occured during its walk. If anyone is now interested to know that the distance between walker's current state and its asymptotic state, then he must tune at a particular dephasing rate keeping all other parameter fixed. Tuning at that particular rate will enable him to know trace norms at all previous times. 
From the trace norm, one can estimate the mixing time in case of $\Gamma_{dep}\rightarrow0$ as 

\begin{eqnarray}
t_{mix}=\frac{log_{e}\left[\frac{\overline{D}[\rho_{dep}^c(t),\rho_{dep}^c(\infty)]-\frac{1}{4\sqrt{N}}}{\frac{\sqrt{(1-\frac{1}{N})^{2}+a^{2}}}{2}-\frac{1}{4\sqrt{N}}}\right]}{log_{e}\left(1-2\Gamma_{dep}\right)}
\end{eqnarray}
\/*
where $\overline{D}_{}\{\rho_{\frac{N}{2}}(t),\rho_{N}(\infty)\}>$ implies the trace norm when the walker has already covered $\frac{N}{2}$ sites. Using Eq() one can also estimate the trace distance between walker's initial state and the state after $t^{th}$ iteration as 
\begin{eqnarray}
<D_{\lambda=\frac{1}{2}}\{\rho_{\frac{N}{2}}(t),\rho(0)\}>=\frac{1}{2}\left\langle ||\rho_{\lambda=\frac{1}{2},\frac{N}{2}}(t)-\rho_{N}(\infty)||\right\rangle\nonumber\\
\geq\frac{1}{4\sqrt{N}}+\frac{\sqrt{\left(1-\frac{1}{N}\right)^{2}+a^{2}}}{2}\left(1-\exp(-2\Gamma_{dep}t)\right)\nonumber\\
\end{eqnarray}
*/
\subsection{Percolation with continuous bit-flip noise}
Now we analyse the mixing time of the quantum walk by considering that bit-flip noise is acting on the qubit continuously. The initial state of the walker is the same as given in Eq.().
The empty graph operation in this case will be 
\begin{equation}
\xi_{0}(\rho_{0}^{c})=
\frac{1}{2}\left(\begin{array}{cc}
1+acos\theta e^{-\Gamma_{bit}t} &  A asin\theta \\
Basin\theta & 1-acos\theta e^{-\Gamma_{bit}t}
\end{array}\right)
\end{equation}
where $A=cos\phi-isin\phi e^{-\Gamma_{bit}t}
$ and $B=cos\phi+isin\phi e^{-\Gamma_{bit}t}$
\/*
After s arbitrary steps the empty graph operation can be given by 
\begin{equation}
\xi_{0,bit}^{s}(\rho_{0}^{c})=
\frac{1}{2}\left(\begin{array}{cc}
1+acos\theta(1-p_{bit})^s &  C asin\theta \}\\
Dasin\theta & 1-acos\theta(1-p_{bit})^s
\end{array}\right)
\end{equation}
where $C=\{cos\phi-isin\phi(1-p_{bit})^s$ and $D=\{cos\phi+isin\phi(1-p_{bit})^s\}$.
In case of continuous bit flipping there is a bit flipping rate $\Gamma_{bit}$ such that one may write
$(1-p_{bit})^{s}=e^{-\Gamma_{bit}t}$. Therefore, for $t\rightarrow\infty$ the empty graph operation can be expressed as
\begin{equation}
\xi_{0,bit}^{\infty}=\frac{1}{2}\left(\begin{array}{cc}
1 & asin\theta sin\phi\\
asin\theta sin\phi & 1
\end{array}\right)
\end{equation}

In case of bit flip noise, we get the trace norm as 
\begin{eqnarray}
2D[\rho_{bit}(m,t,\lambda=\frac{1}{2}),\rho_{bit}(N,\infty)]\nonumber\\
\leq\left[\left\{ P_{0}(m,t)(1+acos\theta e^{-\Gamma_{bit}t_{0}})-\frac{1}{N}\right\} ^{2}+P_{0}^{2}(m,t)a^{2}(cos^{2}\phi+sin^{2}\phi e^{-2\Gamma_{bit}t_{0}})\right]^{\frac{1}{2}}\nonumber\\
+4m\left[\left\{ P_{m}(t)(1+acos\theta e^{-\Gamma_{bit}t})-\frac{1}{N}\right\} ^{2}+P_{m}^{2}(t)a^{2}(cos^{2}\phi+sin^{2}\phi e^{-2\Gamma_{bit}t})\right]^{\frac{1}{2}}\nonumber\\
+\sum_{k=1}^{M-1}4k\left[\left\{ P_{k}(m,t_{k})\left(1+acos\frac{\theta}{2}(1+e^{-3\Gamma_{bit}})e^{-\Gamma_{bit}t_{k}}\right)-\frac{1}{N}\right\} +\frac{1}{4}P_{k}^{2}(m,t_{k})a^{2}sin^{2}\theta sin^{2}\phi\left(1+e^{-3\Gamma_{bit}}\right)e^{-2\Gamma_{bit}t_{k}}\right]^{\frac{1}{2}}
\end{eqnarray}
*/
Using the above equations and proceeding similar to previous calculations we can find out the trace norm 
$D[\rho_{bit}^c(t),\rho_{bit}^c(\infty)]$ at $t=t_{mix}$. For that we assume
\begin{eqnarray}
|P_{0}(t)- P_u|&\simeq&|P_{1}(t)-P_u|\nonumber\\
&\simeq&...................\nonumber\\
&\simeq&|P_{M}(t)-P_u|\simeq\delta_{bit}\nonumber\\
\end{eqnarray}
Now, for $\phi=\frac{\pi}{2}$, the coin state will become maximally mixed in the asymptotic limit. Therefore, by taking $\phi=\frac{\pi}{2}$ and $\lambda=\frac{1}{2}$ we get the average trace norm at $t\simeq t_{mix}$ as
\/*
\begin{eqnarray}
&D_{\phi=\frac{\pi}{2}}\left[\rho_{\lambda=\frac{1}{2}}\left(\frac{N}{2},t_{mix}\right),\rho\left(N,\infty\right)\right]\leq\frac{1}{4}N\delta_{bit}&\nonumber\\
&+2M\delta_{bit}\left(1+\frac{1}{N\delta_{bit}}\right)acos\theta e^{-\Gamma_{bit}t_{mix}}&\nonumber\\
&+M\delta_{bit}\left(1+\frac{1}{N\delta_{bit}}\right)\frac{e^{-(t_{_{mix}}+3)\Gamma_{bit}}\left(1+e^{-3\Gamma_{bit}}\right)\left(1-e^{-\frac{3M}{2}\Gamma_{bit}}\right)}{\left(1-e^{-3\Gamma_{bit}}\right)}&\nonumber\\
&+M\delta_{bit}\left(1+\frac{1}{N\delta_{bit}}\right)^{2}a^{2}\left(1+cos^{2}\theta\right)e^{-2\Gamma_{bit}t_{mix}}&\nonumber\\
&+\frac{M\delta_{bit}}{4}\left(1+\frac{1}{N\delta_{bit}}\right)^{2}\frac{e^{-2(t_{mix}+3)\Gamma_{bit}}\left(1+e^{-3\Gamma_{bit}}\right)^{2}\left(1-e^{-3M\Gamma_{bit}}\right)}{\left(1-e^{-6\Gamma_{bit}}\right)}&\nonumber\\
\end{eqnarray} 
\\

Here we are interested to study that scinario where the coin state becomes maximally mixed in the asymptotic limit. So, keeping $\phi$ fixed at $\frac{\pi}{2}$, we take average over the polar coordinate $\theta$ over $[-\pi,\pi]$ which means
\begin{equation}

\left\langle D_{\phi=\frac{\pi}{2}}(t_{mix})\right\rangle =\frac{1}{2\pi}\int_{-\pi}^{\pi}D\left(\theta,\phi=\frac{\pi}{2},t_{mix}\right)d\theta
\end{equation}
Thus, taking average we get 
*/

\begin{eqnarray}
\overline{D}[\rho_{bit}^c (t),\rho_{bit}^c (\infty)] &\leq&\frac{N\delta_{bit}}{4}\nonumber\\
&&+\left\{\frac{M\delta_{bit}}{2}a^{2}\left(1+\frac{1}{N\delta_{bit}}\right)^{2}\right\}\nonumber\\
&& \times \left\{ 3+f_{bit}(M,\Gamma_{bit})\right\} e^{-2\Gamma_{bit}t_{mix}}\nonumber\\
\end{eqnarray}
where $f_{bit}(M,\Gamma_{bit})=\frac{e^{-6\Gamma_{bit}}\left(1+e^{-3\Gamma_{bit}}\right)^{2}\left(1-e^{-3M\Gamma_{bit}}\right)}{2\left(1-e^{-6\Gamma_{bit}}\right)}$ is a function that depends on the noise rate as well as on lattice size. Now, analytically one can calculate the trace distance of initial and the asymptotic state as 
\begin{equation}
D[\rho_{bit}^c(t),\rho_{bit}^c(\infty)]=
\frac{1}{2}\sqrt{\left\{ \left(1-\frac{1}{N}\right)+acos\theta\right\} ^{2}+a^{2}}
\end{equation}
Using Simpson's $\frac{1}{3}^{rd}$ rule, one can numerically calculate the average trace norm over $\theta\in[-\pi,\pi]$ which gives different values of it against different values of coin parameter $a$. The values are given in table 1.

\begin{center}
\begin{tabular}{||c|c||}
\hline
$\left\langle D_{\phi=\frac{\pi}{2}}\left[\rho\left(t=0,\theta\right),\rho(N,\infty)\right]\right\rangle$ & $a$\\ [2.0ex]
\hline\hline
0.75 & 1.0\\
\hline
0.71 & 0.9\\
\hline
0.67 & 0.8\\
\hline
0.63 & 0.7\\
\hline
0.59 & 0.6\\
\hline
0.56 & 0.5\\
\hline
\end{tabular}
\end{center}

\begin{figure}[t!]
\centering
\includegraphics[height=7cm,width=7cm]{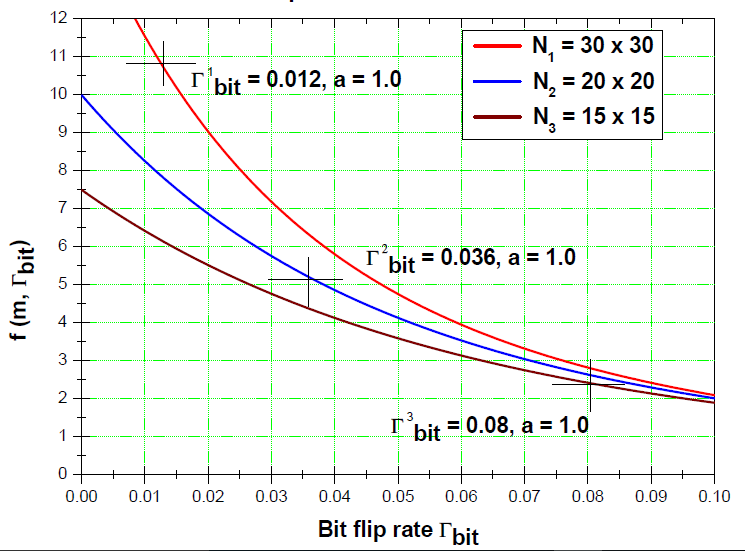}
\caption{(Color on-line) Tunable bit flip rate for three different lattice size. }\label{fig4}
\end{figure}

\begin{figure}[t!]
\centering
\includegraphics[height=7cm,width=7cm]{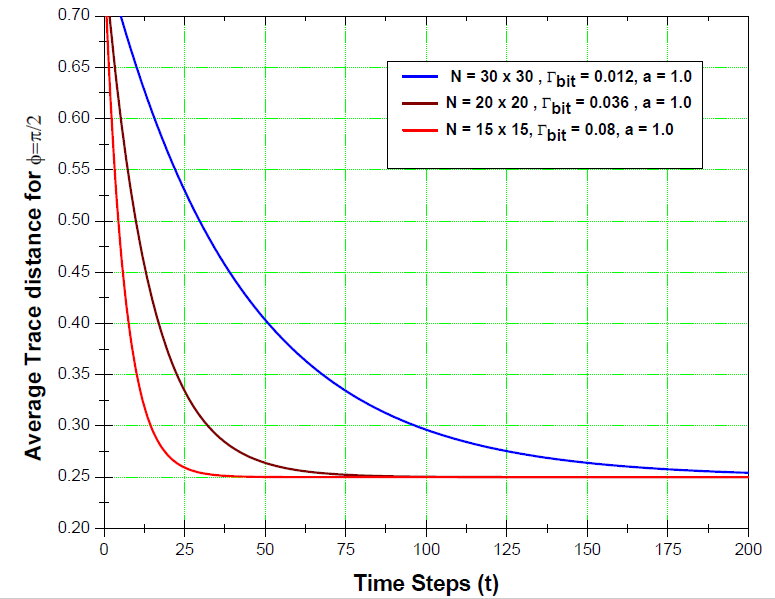}
\caption{(Color on-line) Time evolution of average trace distance ($\overline{D}(t,\delta))$ under continuous bit flip channel. }\label{fig5}
\end{figure}
\paragraph*{}
Now, if we want to calculate the time evolution of trace norm for $0\leq t\leq\infty$, analytically one can say that
\begin{eqnarray}
\overline{D}[\rho_{bit}^c(t),\rho_{bit}^c(\infty)] &=&\overline D[\rho_{bit}^c(t),\rho_{bit}^c(\infty)]\nonumber\\
&&+\overline{D}[\rho_{bit}^c(t=0),\rho_{bit}^c(\infty)] e^{-2\Gamma_{bit}t}\nonumber\\
&&-\overline{D}[\rho_{bit}^c(t),\rho_{bit}^c(\infty)] e^{-2\Gamma_{bit}t}\nonumber\\
\end{eqnarray} 
So, the exponentially converging form of trace norm towards the asympotitc state is possible iff one can say that
\begin{eqnarray}
&\frac{1}{2}M\delta_{bit}\left(1+\frac{1}{N\delta_{bit}}\right)^{2}a^{2}\left\{ 3+f_{bit}(M,\Gamma_{bit})\right\} =&\nonumber\\
&\overline{ D}[\rho_{bit}^c(t=0),\rho_{bit}^c(\infty)] -\overline{D}[\rho_{bit}^c(t),\rho_{bit}^c(\infty)] &\nonumber
\end{eqnarray}
The above equality will satisfy iff we choose $\delta_{bit}\simeq\mathcal{O}\left(\frac{1}{N}\right)$, this value of $\delta_{bit}$ increases the separation from uniform distribution over $(M\times M)$ sublattice. 
Putting the chosen value of $\delta_{bit}$ in the above equation gives 
\[
\frac{9}{4\sqrt{N}}a^{2}\left\{ 3+f_{bit}(M,\Gamma_{bit})\right\} +\frac{1}{4}=\overline{D}[\rho_{bit}^c(t=0),\rho_{bit}^c(\infty)] 
\]
Thus evolution of trace norm simply becomes 
\begin{eqnarray}
\overline D[\rho_{bit}^c(t),\rho_{bit}^c(\infty)]&=&\frac{1}{4}(1-e^{-2\Gamma_{bit}t})\nonumber\\
&&+\overline{D}[\rho_{bit}^c(t=0),\rho_{bit}^c(\infty)]e^{-2\Gamma_{bit}t}\nonumber\\
\end{eqnarray}
The above equation implies an exponential convergence of trace norm towards a fixed number $0.25$. So spreading of walk inside the largest square sublattice remains very far from the uniform distribution. Whereas under continuous dephasing this spreading becomes more and more close to the uniform distribution with increasing size of lattice. 
From the time evolution of trace norm one can estimate roughly the mixing time $t_{mix}^{bit}$ as
\begin{equation}
t_{mix}^{bit}\simeq\frac{log_{e}\left(\frac{\overline D[\rho_{bit}^c(t),\rho_{bit}^c(\infty)] -\frac{1}{4}}{\overline{D}[\rho_{bit}^c(t),\rho_{bit}^c(\infty)] -\frac{1}{4}}\right)}{log_{e}\left(1-2\Gamma_{bit}\right)}
\end{equation}
 
\section{Conclusion}
In this paper we have studied quantum walk over a square lattice in a scenario where dynamic percolation is changing the graph structure randomly and Markovian noise is acting continuously on the walker (qubit). 
We have considered two types of noises, phase damping and bit-flip, and investigated the quantum walk by constructing zone basis on the square lattice. The initial coin state chosen here is a generic qubit state. The amount of mixedness included in the coin state has been parametrized by the variable $a$. It is obvious that the asymptotic state of the walker in $\mathbb{C}^2\otimes\mathbb{C}^N $ will be $\frac{\mathbf{1}_2}{2}\otimes\frac{\mathbf{1}_N}{N}$, which implies an uniform distribution. Our result reconfirms the fact that quantum walk is faster than the classical one. Thus, for $a\rightarrow 1$ the exponential convergence of the trace distance bewteen two probability distributions, position of the walker at any instant of time and the uniform distribution, will be faster and for $a\rightarrow 0$ it will be slower. Thus, chosing the initial coin state from the surface of the Bloch sphere makes the walk fastest even in the presence of percolation. Moreover, it has been shown that the exponential convergence of the trace distance bewteen the two probability distributions  depends on the rate($\Gamma_{dep}$ or $\Gamma_{bit}$) of Markovian noise. Under dephasing noise the probability distribution of walker's position becomes close to the uniform distribution with the closeness factor $\delta_{dep}\equiv O(N^{-\frac{3}{2}})$.  Whereas, in case of bit flip noise the closeness parameter $\delta_{bit}\equiv O(N^{\frac{1}{n}})$ implies that bit flip noise resists the walk to spread uniformly. Finally, this work can be extended for higher dimensions with multiple walker.

\section*{Acknowledgement}
A. G would like to thank Bose Institute, Kolkata, for the financial support. P. D thanks Department of Science and Technology, Govt. of India, for the financial support and Bose Institute for the research facilities.

\end{document}